# High-Contrast Interferometric Imaging of Single-Molecule Dynamics on Optical Fibers

Guifeng Li[1], Chaoyang Gong[1,*]

[1] Key Laboratory of Optoelectronic Technology and Systems (Ministry of Education of China), School of Optoelectronic Engineering, Chongqing University, Chongqing 400044, China.

Correspondence Email: cygong@cqu.edu.cn

**Abstract**

Single-molecule detection enables direct observation of individual biomolecular events, providing mechanistic insights into biological processes and offering a powerful tool for disease diagnostics. However, the fundamental scale mismatch between optical wavelengths and molecules restricts the application of label-free techniques, leading to poor signal-to-noise (SNR) performance. Here, we propose a high-contrast, label-free approach based on interferometric imaging, utilizing the strong evanescent field supported on a microfiber surface to provide near-field illumination. We observed unique interference patterns generated by in-plane scattering from natural defects, which enabled high-contrast detection of localized phase changes induced by single molecules. The results indicate an approximately 38 dB enhancement in SNR over the conventional fluorescence methods, without employing any plasmonic or microcavity-based amplification techniques. This approach was further applied to track molecular dynamics, capturing both conformational transition and binding behaviors of individual protein molecules. Meanwhile, the stimulus–response of single molecules to acoustic waves was investigated, demonstrating the ultimate miniaturization of an acoustic sensor at the single-molecule scale. By enabling direct observation of molecular dynamics and mechanical responses at the single-molecule level, this approach provides a versatile platform for probing fundamental biological processes and developing ultra-sensitive biosensors. Moreover, this approach lays the foundation for coupling optical and acoustic waves at the molecular scale, opening new avenues for next-generation single-molecule diagnostics and precision biophysics studies.

# 1. Introduction

Investigating molecular dynamics at the single-molecule level provides mechanistic insights into biological processes and plays a crucial role in drug discovery and the detection of disease biomarkers (1-3). Following the first optical detection of single molecules via absorption (4), subsequent approaches have predominantly relied on fluorescence emission(5-7), which enhances signal-to-noise ratio (SNR) by shifting emitted photons to lower energy levels and thereby suppressing background signals. By incorporating the Förster resonance energy transfer (FRET) effect, fluorescence-based methods have become a powerful tool for probing single-molecule dynamics (8, 9). Despite their high sensitivity and SNR, fluorescence-based single-molecule techniques are fundamentally limited by the photophysical properties of fluorophores, including optical saturation, photobleaching, and photoblinking (10-12). In addition, the attachment of fluorescent labels can perturb the native properties and dynamics of the molecules under investigation, potentially compromising the accuracy of the observed behavior (13-16).

Label-free technologies track molecular dynamics without fluorescent tags, offering a detailed view of molecular behavior and properties in their native states (17, 18). However, limited by the mismatch between optical wavelengths and molecular dimensions, most existing label-free approaches suffer from low SNR. Although optical microcavities enhance light–matter interactions by confining photons within a small volume, spectral interrogation methods provide the overall phase changes on the cavities, thus compromising the SNR (19-24). Plasmonic resonators provide localized spectral or phase information of single molecules. However, significant absorption losses of nanoparticles reduce the Q-factor and broaden the resonance linewidth (25-33). Compared with the spectroscopic methods, label-free imaging technologies provide spatial information that isolates the weak signal of a single molecule from the surrounding background, thereby enabling direct visualization of molecular interactions, conformational transitions, and binding events with high sensitivity (34-37). Photothermal microscopy employs the thermal lens effect to indirectly measure the absorption of a heating beam by a nanoscale sample, but the weak refractive index modulation induced by nanoscale absorbers generates only a small phase shift in the probe beam that is easily buried by background fluctuations (38, 39). Interferometric scattering microscopy employs the interference between out-of-plane scattered light and reflected reference light to resolve the slight phase change induced by single molecules, but the strong reflection introduces a large background that reduces image contrast (40-42). Plasmonic scattering microscopy exploits the strong near-field enhancement of metallic nanostructures to boost the scattering signal from single molecules, but the accompanying photothermal effect restricts the maximum excitation power and limits the achievable SNR (43-45).

Here, we propose an interferometric imaging approach that enables high-contrast visualization of analytes down to the single protein level. As illustrated in Fig. 1a, a microfiber was employed for near-field illumination of protein molecules. Natural defects on the fiber surface scatter a fraction of the light, and the scattered light interferes to form a unique interference pattern (Fig. 1b). Thanks to the small diameter of the microfiber, a strong evanescent field was supported on the solid-liquid surface, enabling a relatively strong scattering (Fig. 1c). Meanwhile, the intensity balance between the in-plane scattered fields ensures a high-contrast interference pattern that is independent of the intrinsic scattering cross section of individual molecules (Fig. 1d). Our findings reveal that the proposed interferometric imaging approach achieves an approximately 38 dB enhancement in SNR over conventional fluorescence methods, without employing any plasmonic or microcavity-based amplification techniques (Figs. 1e to 1g). Since no photobleaching effect was involved, the proposed method enables real-time monitoring of molecular dynamics. This technology was employed to monitor the conformation transition and binding dynamics of a single protein molecule. Moreover, the mechanical response of individual protein molecules to acoustic wave was demonstrated, potentially revealing the relationship between molecular structure, dynamics, and external forces. Our label-free interferometric imaging method provides a powerful platform for real-time monitoring of molecular dynamics and opens a door for probing the mechanistic responses of individual biomolecules to external stimulation.

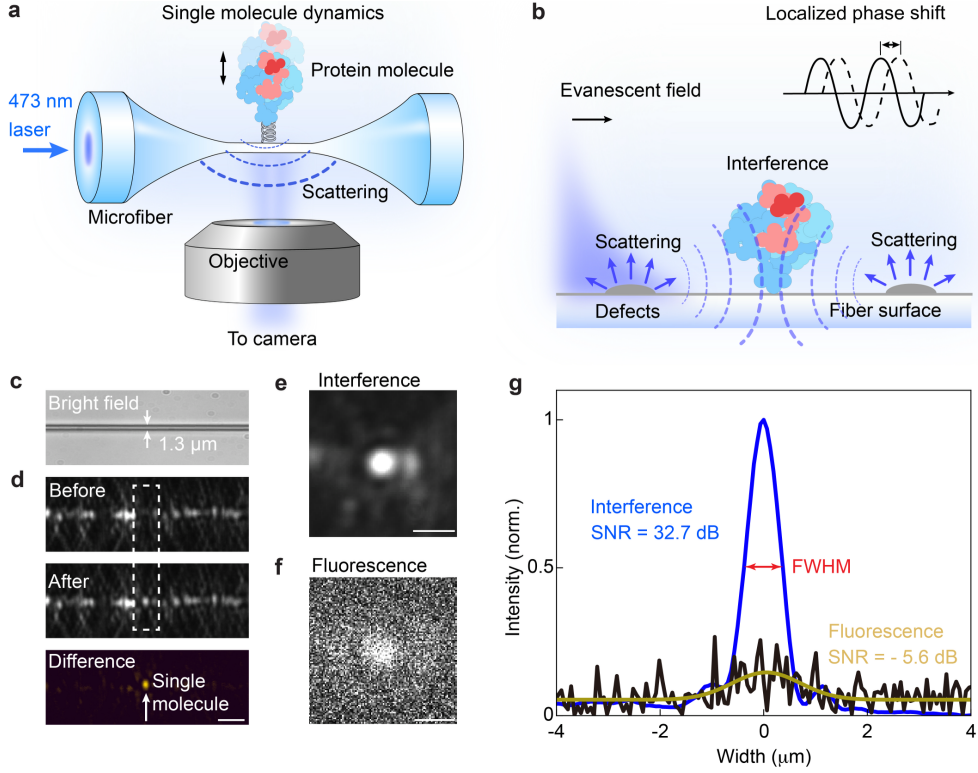

**Figure 1 Conceptual illustration of single-molecule visualization on an optical fiber.** (**a**) Experimental setup. (**b**) Mechanism of single-molecule detection. (**c**) Microscopic image of a microfiber. (**d**) Observation of a single-molecule binding event through interference image. Scale bar: 5 μm. (**e,f**) Comparison of interference (**e**) and fluorescence (**f**) images. Scale bar: 2 μm. (**g**) Comparison of signal-to-noise ratio.

## 2. Results

### 2.1 Theory of interferometric imaging

The microfiber can be regarded as a one-dimensional waveguide. The scattering electric field of the $j$th scatter under incident field $E(x)$ can be written as $F_j = f_j(x) \cdot E(x) \otimes h_c(x)$. Here, $h_c(x)$ is the coherent point spread function and $\otimes$ denotes convolution. The observed scattering image is an interference pattern, which is the coherence superposition of the scattering electric field of each scatterer. Therefore, the observed image can be written as

$$I(x) = \left|\sum_j F_j\right|^2 = \sum_j |F_j|^2 + 2\sum_{j \neq k} |F_j|^2 \cdot |F_k|^2 \cdot cos(\Delta\varphi_{jk}) \quad (1)$$

where $\Delta\varphi_{jk} = \varphi_j - \varphi_k$ denotes the phase difference between the scattering fields of the *j*th and *k*th scatterers. The first term represents the incoherent image of the scatterers. The second term is the coherent image, which is dependent on the phase difference between two scattering fields. The binding of single molecules on the fiber surface induces a localized phase change, thus changing the interference pattern (46).

## 2.2 Interferometric imaging of a single molecule

Bovine serum albumin (BSA) was employed as the model protein to validate the label-free visualization of single-molecule binding events on an optical fiber. The experimental setup is shown in Supplementary Fig. S1. A microfiber with a diameter of approximately 1.3 μm was used as the sensing element (Fig. 1c), and its scattering signal was collected using an inverted microscope. As shown in Fig. 1d, an interference pattern was observed due to the in-plane scattering of light. The Differential images reveal multiple individual bright spots, corresponding to BSA molecules adsorbed onto the fiber surface via non-specific interactions (Fig. 2a). Analysis of the extracted intensity distribution along the optical fiber indicates that molecular binding occurs randomly over time (Fig. 2b). We have to note that the observed bright dots correspond to the Airy patterns, which cannot reveal the real size of molecules (See Supplementary Materials for more details).

Real-time monitoring of BSA adsorption on the microfiber was conducted at different concentrations. The binding kinetic curve was achieved by plotting the number of BSA molecules binding to the surface versus time (Fig. 2c). Once the microfiber was immersed in the solution, the number of binding events increased linearly at first and gradually approached saturation after 337 s. The saturation results from the depletion of molecules and binding sites, leading to a reduced binding probability on the fiber surface. Meanwhile, fewer binding events were observed in lower concentration, which is caused by the decreased molecular collision frequency on the microfiber surface. As shown in Fig. 2d, the number of events shows a good linearity with BSA concentration. Surprisingly, binding events could be observed at a minimum concentration of 0.05 aM, corresponding to approximately 15 molecules in 500 μL aqueous solution. As a control experiment, no binding event was observed without protein molecules (Supplementary Fig. S2). This extremely high sensitivity approaching the physical limit demonstrates strong potential for quantitative analysis at ultra-low concentrations and paves the way for early biomarker detection.

To further validate our observations, conventional fluorescence technology was employed to demonstrate that the observed scattering signal corresponds to single-molecule binding events. The BSA molecules labeled with Rhodamine B (RhB) was used for the experiment, which exhibits a peak fluorescence emission at 593 nm under 520 nm excitation (See Methods and Supplementary Fig. S3 for more details). When the BSA-RhB molecules adsorbed onto the fiber surface, the scattering signal was transmitted by a dichroic mirror while the fluorescence signal was reflected (Fig. 2e). This configuration enables simultaneous monitoring of scattering and fluorescence signals (Figure 1e and f). As shown in Fig. 2f, the fluorescence signal started to appear at 1043 s, and the scattering signal arose simultaneously. This result further confirms that the scattering signal reveals the single molecule binding event successfully.

Compared with the conventional fluorescence method, the label-free feature of the scattering image eliminates the influence of photobleaching effect, enabling long-term monitoring of molecular interactions. As shown in Fig. 2f, the intensity fluctuations of the scattering images reveal the molecular dynamics, whereas the fluorescence signal rapidly photobleached after 63 seconds of photon illumination. These results indicate that our label-free method provides real-time insights into molecular dynamics at the single-molecule level.

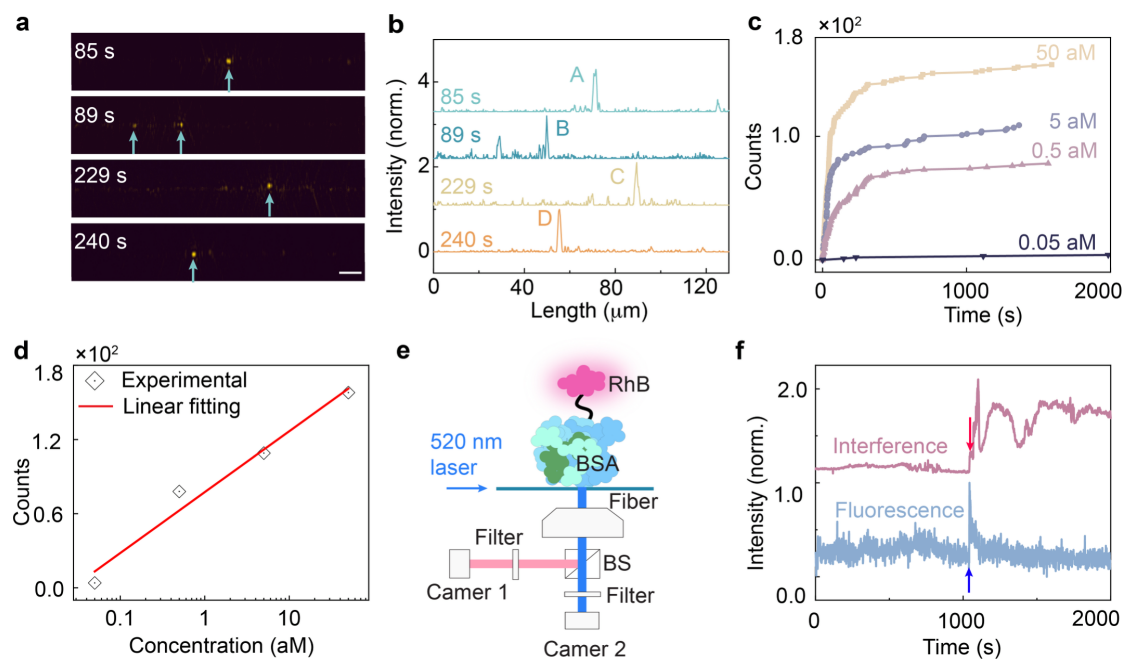

**Figure 2 Single-molecule detection.** (**a,b**) Differential images of interference patterns (**a**) and the corresponding spatial distributions (**b**) obtained at various times. False color was applied to enhance visualization. Scale bar: 10 μm. (**c**) Evolution of the number of BSA binding events recorded at various concentrations. (**d**) The number of BSA binding events as a function of concentration. Data are extracted from (**c**) at 1984 s. (**e**) Illustration of the dual-channel imaging system. (**f**) Comparison of the time-resolved intensity evolution between interference and fluorescence images of a single BSA adsorption. The arrows indicate the onset time of signals.

## 2.3 Real-time monitoring of conformation transition

Then, we employed the scattering image to reveal the conformational dynamics, which is essential for understanding various biochemical processes, including molecular recognition (47, 48), cell signaling (49, 50), and transcriptional regulation (51, 52). As illustrated in Fig. 3a, BSA molecules undergoes pH-dependent conformational transition due to changes in intramolecular charge distribution and electrostatic interactions (53). At neutral pH, BSA molecules maintain their native conformation. As the pH decreases, increased protonation raises the net positive charge and electrostatic repulsion, thereby disrupting native tertiary contacts and inducing progressive partial unfolding transitions. As shown in Fig. 3b, BSA molecules under acidic conditions exhibit a faster binding rate due to the protonation of acidic residues. Meanwhile, the acidic environment triggers protein unfolding, which increases the effective distance between the molecule and the fiber surface, resulting in a lower scattering intensity (Fig. 3c) (54).

Direct visualization of the unfolding process was further achieved by recording the real-time interference images at various pH conditions. Figure 3d shows that the molecular scattering signal remains constant at neutral pH. When an acidic condition was applied, the scattering intensity of a single molecule continuously decreased, corresponding to a dynamic unfolding process. The dynamic scattering intensity under various pH conditions is compared in Figure 3e, where a more pronounced decrease at lower pH suggests an increased rate of conformational transition. These findings establish our scattering-based imaging technique as a powerful tool for real-time monitoring of structural dynamics at the single-molecule level, offering unprecedented sensitivity in capturing complex molecular behaviors.

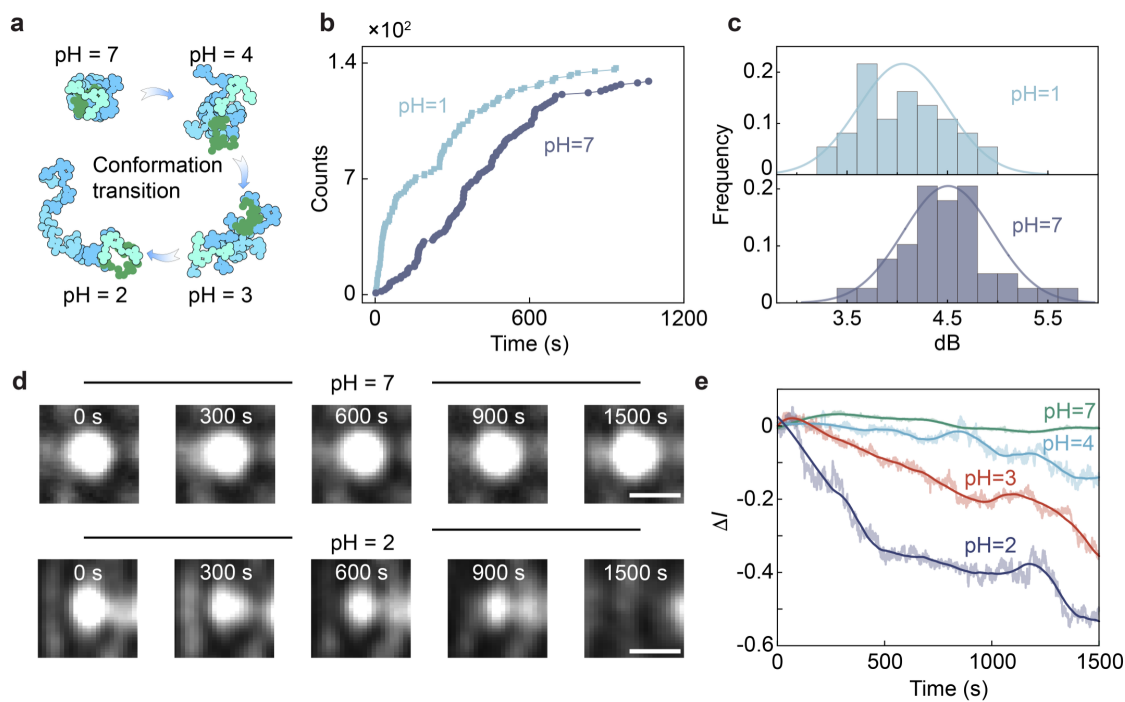

**Figure 3 Conformation transition.** (**a**) Illustration of conformational transitions of BSA under different pH values. (**b,c**) Binding dynamics curves (**b**) and statistical distribution of scattering intensity (**c**) recorded at different pH conditions. (**d**) Time-resolved imaging of single BSA molecules under different pH conditions. Scale bar: 2

μm. (**e**) Temporal evolution of intensity difference (Δ*I*) at various pH conditions.

**2.4 Molecular binding dynamics**

As a conceptual illustration, we investigated the binding dynamics of single molecules. Owing to the real-time monitoring capability of scattering imaging, three distinct types of molecular behavior were observed, including "binding", "unbinding", and "dancing" (Fig. 4a-c left). Due to variations in molecular orientation and position after binding to the fiber surface, the attractive forces exerted by the surface differ in each binding event. A sufficiently strong binding force stabilizes the attachment of molecules to the fiber surface ("binding"), while a relatively weak force leads to their detachment ("unbinding"). In contrast, when the binding force is moderate, the molecules neither remain fully attached nor completely unattached. Instead, they undergo continuous fluctuations, transiently binding and releasing from the surface, resulting in a dynamic equilibrium state, which we term "dancing". As shown in Figs. 4a-c (right) and 4d, the three types of dynamic behaviors are characterized by distinct features, corresponding to a step-like pattern, a square-wave pattern, and random fluctuations in scattering intensity, respectively. These results are consistent with the reported molecular binding dynamics revealed by plasmonic scattering microscopy, further confirming the reliability of our method (55). We compared the intensity fluctuations of BSA and IgY molecules exhibiting "dancing" behavior in Fig. 4e and 4g. The result shows that the intensity fluctuation of IgY molecules is much stronger than BSA, resulting in a broader statistical distribution curve (Fig. 4f and h). We think this result arises from the larger molecular size of IgY, which induces stronger localized phase variations during the "dancing" process. We envision that such "dancing" behavior could serve as a potential indicator for revealing the heterogeneity of individual molecules.

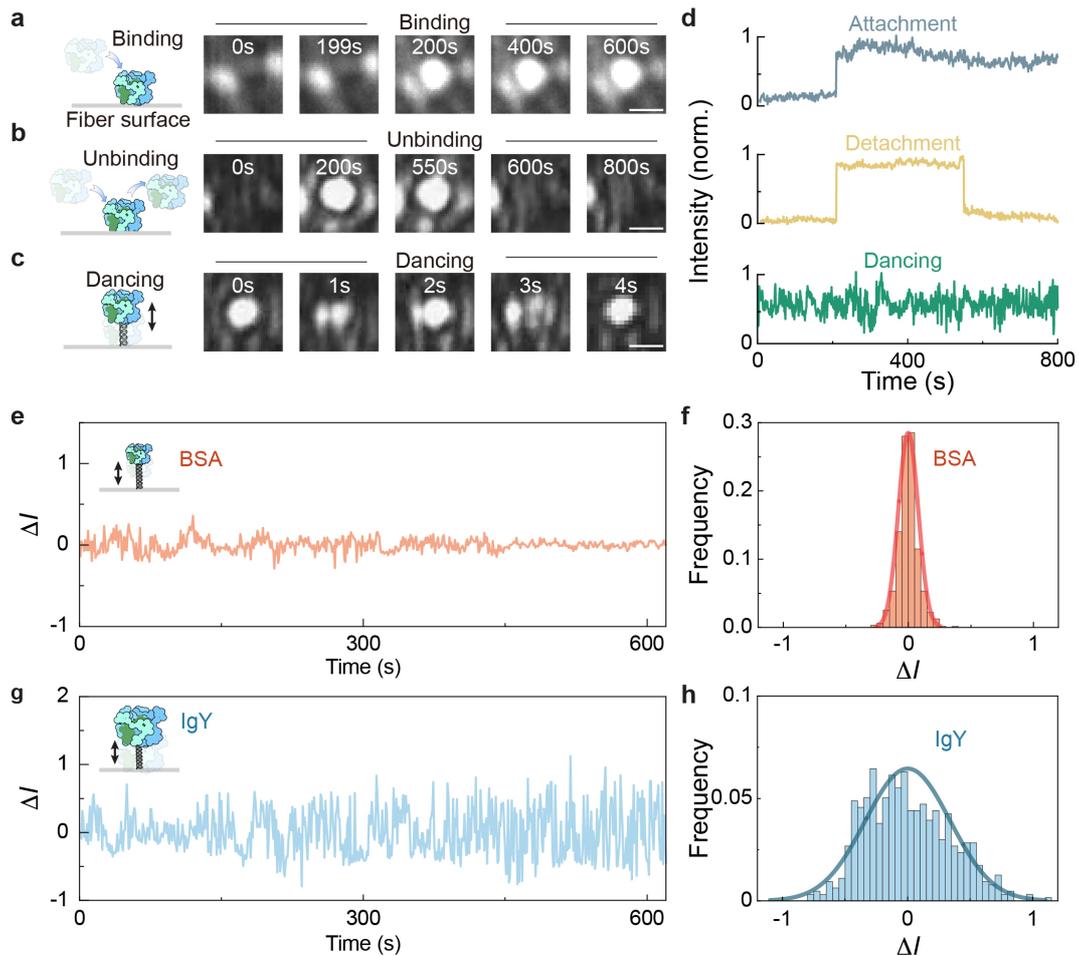

**Figure 4 Binding dynamics of a single BSA molecule.** (**a,b,c**) Conceptual illustrations (**left**) and time-resolved images (**right**) of the dynamic binding behavior of a single molecule, including "binding" (**a**), "unbinding" (**b**), and "dancing" (**c**). Scale bar: 2 μm. (**d**) Intensity signals of single-molecule binding dynamics. (**e-h**) Comparison of the temporal evolution of $\Delta I$ for dancing BSA and IgY molecules (**e,g**) and the corresponding statistical distributions (**f,h**).

## 2.5 Response of a single molecule to acoustic stimulation

The stimulus-response of single molecules to the acoustic wave was investigated. As shown in Fig. 5a, the acoustic wave exerts forces on the molecules, thereby modulating their binding dynamics. Molecules near the optical fiber were accelerated by acoustic excitation, and some were driven onto the fiber surface, as indicated by a stepwise increase in scattering intensity (Fig. 5b). Meanwhile, molecules adsorbed on the fiber surface exhibited unbinding behavior in response to the acoustic perturbation (Fig. 5c).

In addition, we found that the "dancing" behavior of a single molecule can be modulated by the acoustic wave. As shown in Fig. 5d, the real-time intensity exhibits no significant changes following acoustic wave excitation, likely due to strong noise arising from Brownian motion. However, after passing through a narrowband filter, the signal shows a more pronounced intensity fluctuations upon acoustic wave application. As shown in Fig. 5e, the fast Fourier transform (FFT) of the temporal intensity reveals distinct frequency components at different excitation frequencies. As the acoustic wave excitation frequency is varied, the corresponding frequency peaks shift accordingly, and the experimental results at different excitation frequencies agree well with the theoretical calculations (Fig. 5f). In contrast, the frequency peaks disappear when the acoustic wave is turned off. Furthermore, the detected acoustic wave intensity increases with the increasing driving voltage (Fig. 5g). Our work investigates the stimulus–response of molecular dynamics to acoustic stimulation, which may represent an ultimate limit of miniaturized acoustic wave sensing.

The obtained frequency is about three orders of magnitude lower than the acoustic wave excitation (503 Hz, 503.5 Hz, 504 Hz and 504.5 Hz), which is due to the under sampling effect (Supplementary Fig. S4)(56, 57). The camera samples the continuous waveform of acoustic wave at a low frame rate (20 fps), producing a sequence of discrete values that constitute a lower-frequency waveform. We carried out theoretical simulation to further illustrate the under sampling effect. As illustrated in Fig. 5h, although the under sampling effect distorts the acoustic wave signal, the frequency shift of the under sampling signal remains consistent with that of the original signal, which can be further employed to quantify the acoustic signal.

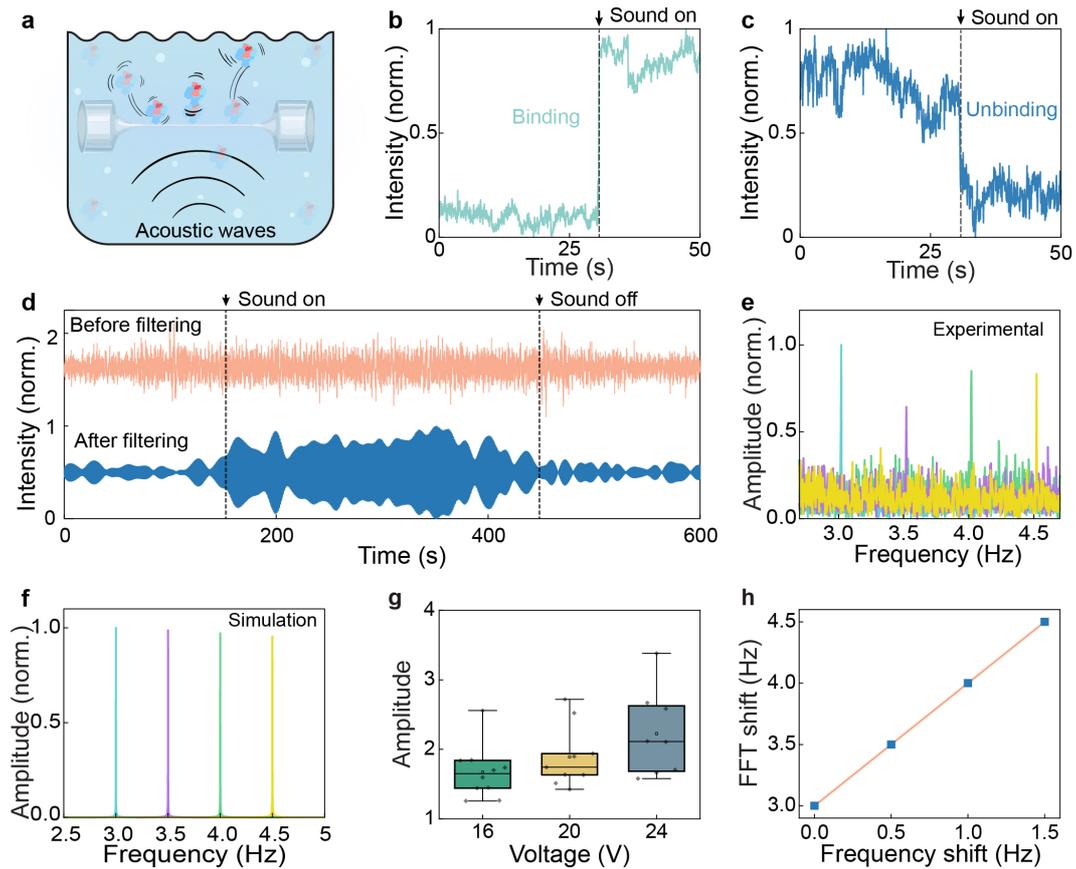

**Figure 5 Dynamic response to acoustic stimulation.** (**a**) Illustration of single-molecule response behavior under acoustic wave stimulation. (**b-c**) Time-resolved image intensity of individual molecules at fixed positions (**b**) binding behavior (**c**) unbinding behavior. (**d**) Comparison of temporal intensity traces of dancing molecules before and after band-pass filtering. (**e-f**) Different acoustic excitation frequencies (**e**) FFT transforms of intensity traces (**f**) and their corresponding simulation comparisons under. (**g**) Relationship between FFT peaks and driving voltage under a fixed acoustic frequency. Error bars represent standard deviations of ~10 measurements. (**h**) Relationship between the frequency shift of undersampled signals and that of original signals.

## 3. Discussion

We proposed an interferometric imaging method enabling high-contrast visualization at the single protein level. The evanescent field supported by microfibers was used to probe single-molecule dynamics. We achieved real-time monitoring of single-molecule dynamics, monitored the conformational transitions and binding kinetics of individual protein molecules, and demonstrated the mechanical response of single protein molecules to acoustic wave.

Here, we would like to point out the significance of this study. First, the proposed label-free interferometric imaging method serves as a new platform for real-time monitoring of molecular dynamics. Although fluorescence-based techniques have become powerful tools for probing single-molecule dynamics, most of them are limited by the photophysical properties of fluorophores—this not only impairs the accuracy of observed molecular behaviors but also restricts the duration of long-term observation. While label-free techniques can capture the behaviors and characteristics of molecules in their native states, most existing label-free methods suffer from a low signal-to-noise ratio (SNR) due to the mismatch between optical wavelengths and molecular sizes. The label-free interferometric imaging method we proposed provides a powerful platform for enhancing the SNR of single-molecule detection and enabling long-term real-time monitoring of molecular dynamics, which is expected to unlock new possibilities in long-term single-molecule dynamics monitoring and high-SNR imaging. Second, this study also demonstrates the stimulus response of individual protein molecules to acoustic waves, which will offer more details and new opportunities for revealing the relationship between molecular structure, dynamics, and external forces.

Finally, we discuss the advantages and disadvantages of label-free interferometric imaging in probing single-molecule dynamics. Compared to conventional fluorescence methods and other label-free approaches, the advantages of our method are twofold. First, interferometric imaging enables long-term monitoring of molecular behaviors and characteristics without the need for additional labels. Second, the molecular dynamics signals acquired using interferometric imaging possess an ultra-high signal-to-noise ratio (SNR), which helps us explore more details in single-molecule dynamics. However, the temporal resolution of obtaining molecular dynamics information using our proposed method is limited by the frame rate of the signal acquisition and imaging system. Capturing more details of single-molecule dynamics requires the cooperation of cameras with higher imaging frame rates. Nonetheless, interferometric imaging, as a novel single-molecule detection technique, holds significant importance, particularly in fields demanding molecular detection with higher SNR and sensitivity.

## 4. Materials and methods

### 4.1 Experimental setup

The details of the experimental setup are illustrated in Supplementary Fig. S1. An inverted microscope (Chongqing Coic Industrial Co., Ltd, DSZ5000X) mounted with a 100× (NA = 0.8) objective was employed for scattering signal collection. Laser emission (473 nm) from a laser diode (Changchun New Industries Optoelectronics Technology Co., Ltd., MBL-FN-473-100mW) was used to excite the evanescent wave, with approximately 4.5 mW coupled into the microfiber. The scattering signal was sent to a charge-coupled device camera (CCD) (Princeton Instruments, PIXIS 1024) and a spectrograph system (Zolix, Omni-λ3004i) for image and hyperspectral image recording, respectively. The scattering images were recorded by the CCD at 1 fps.

### 4.2 Fabrication of microfiber

The microfiber was fabricated using the melt-drawing method. The single-mode optical fiber (Corning, SMF-28e) was employed as the starting material, and a melting taper machine (OC-2010) was utilized for the drawing process. During fabrication, a hydrogen gas flow rate of 180 sccm was maintained. The drawing length was set to 2.9 cm, with a drawing speed of 0.25 mm/s throughout the process. In order to obtain a clean microfiber surface and enhance protein affinity, we used a plasma cleaner (PLUTOVAC, PLUTO-T) to perform plasma treatment for 60 s, aiming to achieve hydroxylation of the surface.

### 4.3 Hyperspectral imaging

Hyperspectral images were recorded using the spectrograph system. The microfiber was oriented parallel to the entrance slit (~ 10 μm). A broadband supercontinuum laser (Wuhan Yangtze Soton Laser Co., Ltd., SC-OEM) was used as the light source instead of a laser diode. The visible portion of the supercontinuum laser (400-750 nm), selected using a broadband filter, was then coupled into the microfiber. The scattering signal was dispersed by a grating (150 lines.mm$^{-1}$) according to wavelength, allowing different spectral components to be distinguished at various locations on the camera. An electron-multiplying CCD (EMCCD) (Andor, DL-604M-OEM) mounted on the output port of the spectrograph system was employed to record the hyperspectral images.

### 4.4 Single molecule detection of BSA

The stock BSA solution was prepared by dissolving BSA powder (Aladdin, B265994) in phosphate-buffered saline (PBS) solution (pH = 7.4). The BSA solutions with concentrations ranging from 0.05 to 50 aM was freshly prepared with serial dilutions of the stock solution using PBS. The silanized microfiber was immersed in the BSA solution (Supplementary Fig. S1), and the time-resolved scattering images were recorded by the CCD camera.

### 4.5 Real time monitoring of molecular conformational transitions

Figure 3a shows the conceptual illustration of the conformational transition of BSA molecules at different pH values. The BSA stock solution with a concentration of 50aM was prepared by dissolving BSA powder in PBS with pH=7. The micro-optical fiber was immersed in the protein solution to adsorb protein molecules. When the protein adsorption on the surface of the micro-optical fiber was stable, glacial acetic acid (aladdin, A116166) diluted by different multiples was added to the protein solution to adjust the protein solution to acidic (pH=1, 2, 3, 4). After adjusting the protein solution, the scientific complementary metal-oxide-semiconductor camera was used to record the real-time change of the scattering intensity of the molecules adsorbed at pH=7 with time after the pH was changed. The camera records the image with a frame rate of 1 fps.

### 4.6 Actuating the Molecule with acoustic wave

A piezoelectric transducer attached on the glass slide was used to generate acoustic wave. A sine wave with an amplitude adjusted to values of 16 V, 20 V, and 24 V, and a frequency adjusted to values of 503 Hz, 503.5 Hz, 504 Hz, and 504.5 Hz，generated by a function generator, was used to drive the piezoelectric transducer.

**Acknowledgments**

This work is supported by the National Natural Science Foundation of China (Grant No. 62375030); the Fundamental Research Funds for the Central Universities (Grant No. 2024CDJYXTD-004).

**Data availability**

All data are available within the Article and Supplementary Files, or available from the corresponding authors on reasonable request.

**Conflict of interest**

All the authors declare no conflict of interests.